\documentclass[conference]{IEEEtran}
\ifCLASSINFOpdf
\else
\fi
\usepackage[cmex10]{amsmath}
\ifCLASSOPTIONcompsoc
  \usepackage[caption=false,font=normalsize,labelfont=sf,textfont=sf]{subfig}
\else
  \usepackage[caption=false,font=footnotesize]{subfig}
\fi
\usepackage{fixltx2e}
\newcommand{\matr}[1]{\mathbf{#1}}

\DeclareMathOperator{\hermit}{H}
\DeclareMathOperator{\expec}{E}
\DeclareMathOperator{\diff}{d}
\renewcommand{\Re}{\operatorname{Re}}
\renewcommand{\Im}{\operatorname{Im}}

\usepackage{bm}
\usepackage{stfloats}

\usepackage{pgfplots}
\usetikzlibrary{shapes}
\pgfplotsset{compat=newest}
\newlength\figureheight	
\newlength\figurewidth

\newcommand{\columnplot}{
\setlength\figureheight{0.24\textwidth}
\setlength\figurewidth{0.37\textwidth}
}	

\newcommand{\columnplottwo}{
\setlength\figureheight{0.24\textwidth}
\setlength\figurewidth{0.37\textwidth}
}

\usepackage[exponent-product=\cdot]{siunitx}

\begin{document}
%
\title{Spatial Oversampling in LOS MIMO Systems with 1-Bit Quantization at the Receiver}

\author{\IEEEauthorblockN{Tim H{\"a}lsig and Berthold Lankl}
\IEEEauthorblockA{Institute for Communications Engineering\\
Universit{\"a}t der Bundeswehr M{\"u}nchen, Germany\\
Email: tim.haelsig@unibw.de}
}


%


\maketitle

\begin{abstract}
In this paper we investigate the achievable rate of LOS MIMO systems that use 1-bit quantization and spatial oversampling at the receiver. We propose that by using additional antennas at the receiver, the loss incurred due to the strong limitation of the quantization can be decreased. Mutual information results show that considerable rate gains can be achieved depending on the number and arrangement of the antennas. In some of the cases, even the full available rate from the transmitter can be attained. Furthermore, the results also reveal that two-dimensional antenna arrays can benefit more from spatial oversampling than one-dimensional arrays, when using 1-bit quantization in the LOS MIMO scenario.
\end{abstract}


\let\thefootnote\relax\footnotetext{The authors would like to thank M.~El~Chamaa and X.~Song for insightful discussions on the topic. This work was supported in part by the German Research Foundation (DFG) in the framework of priority program SPP~1655 "Wireless Ultra High Data Rate Communication for Mobile Internet Access".}

%
\IEEEpeerreviewmaketitle

\section{Introduction}

To achieve the projected high data rates for forthcoming wireless communications systems, MIMO techniques, with their high spectral efficiencies, and millimeter wave carrier frequencies, with their high available bandwidth, are two of the main focus points. High bandwidths typically require high sampling rates, for which the conventional high-resolution analog-to-digital~converter~(ADC) becomes one of the main power consumers in the system. The usage of MIMO techniques only enhances this effect due to the fact that multiple ADCs are utilized. Two approaches are taken in order to alleviate this problem. One is the consideration of hybrid signal processing approaches, to reduce the number of ADCs, e.g., hybrid beamforming, or to reduce the required resolution of the ADCs, e.g., analog equalization \cite{Song2016}. The other approach looks at the applicability of low-resolution ADCs, down to the extremest case of only one bit of amplitude resolution, as such converters can be built to be very power efficient. Systems operating at millimeter wave carrier frequencies may also make use of line-of-sight~(LOS) MIMO designs achieving high spectral efficiencies through, for example, spatial multiplexing, by separating the transmit and receive antennas such that the spherical wavefronts generate orthogonal channels \cite{Torkildson2011,Halsig2015}. 

In the literature low-bit, and specifically 1-bit, quantization has been investigated for SISO and MIMO systems. For SISO, it was shown in \cite{Krone2012,Halsig2014,Landau2015} that oversampling a signal in the temporal domain can deliver significant information rate improvements, when using 1-bit ADCs. In \cite{Mezghani2007} the authors showed that the capacity loss with 1-bit quantization for Rayleigh fading MIMO channels is small at low SNRs and that QPSK is the best of the investigated input distributions, if there is no channel knowledge at the transmitter. A capacity analysis with channel knowledge at the transmitter was performed in \cite{Mo2015}. The authors provide capacity bounds, for finite and infinite SNR, and a method to find suitable input distributions for this case. In \cite{Corey2016} spatial and temporal sigma-delta signal acquisition for beamforming arrays with coarse quantization is considered. By oversampling the signal in both time and space, and using a sigma-delta structure, a noise shaping is achieved and the performance can be improved. A more complete treatise on parallel spatial sigma-delta receiver design with low precision quantization for channels with a dominant path can be found in \cite{Palguna2016}. In \cite{Ugurlu2016} a strategy to improve the decoding success in low-quantized multiantenna systems is presented. The idea is to add a specific amount of pseudo-random white Gaussian noise in order to provide a dithering effect across the receive antennas that allows for a better symbol recovery after quantization.

In this paper we show that if 1-bit quantization is used at the receiver side of LOS MIMO systems, adding additional spatial sampling points improves the achievable rate depending on the antenna array setup and the input alphabet. Therefore, just as in the temporal domain, some of the information that would otherwise be lost due to the quantization can be recovered by oversampling the received signal spatially. 
In contrast to other channel types \cite{Jacobsson2015}, the LOS channel is very specific in that the spatial multiplexing gain is often limited by the maximum allowed array sizes rather than the number of antennas \cite{Torkildson2011}. As a consequence the loss incurred by the low resolution quantization will be less severe compared to other system types, even at higher SNR levels.

Consider $(\cdot)^T$ and $(\cdot)^H$ to denote transpose and conjugate transpose, respectively. Boldface small letters, e.g., $\matr{x}$, are used for vectors while boldface capital letters, e.g., $\matr{X}$, are used for matrices. Furthermore, $\matr{I}_N$ is the $N\times N$ identity matrix and $\expec [\cdot]$ is the expected value.

\section{System Model}

Assume the narrowband discrete-time received vector of a MIMO system in equivalent complex baseband as
\begin{equation}
\matr{y} = \matr{H}\matr{x} + \matr{n} \text{,}
\end{equation}
where $\matr{y}=\left[y_{1},\ldots,y_{M}\right]^T$, $\matr{x}=\left[x_{1},\ldots,x_{N}\right]^T$ and $\matr{n}=\left[n_{1},\ldots,n_{M}\right]^T$, with $n=1\ldots N$ and $m=1\ldots M$ being the index and number of transmit and receive antennas, respectively. Considering one independent front end for each receiving antenna and assuming that they are the main source of noise, the noise samples are uncorrelated and have a complex Gaussian distribution of $\matr{n}\sim\mathcal{C}\mathcal{N}(0,\sigma^2\matr{I}_{M})$.

The 1-bit quantized receive vector is given as 
\begin{equation}
\matr{y}_Q = \matr{Q}_1\left[\matr{y}\right] \text{,}
\end{equation}
where $\matr{y}_Q=\left[y_{Q,1},\ldots,y_{Q,M}\right]^T$ and where $\matr{Q}_1[\cdot]$ denotes the 1-bit quantization of the whole system, which is given per receive antenna as
\begin{equation}
y_{Q,m} = Q_1\left[\Re\left\{y_m\right\}\right] + j\cdot Q_1\left[\Im\left\{y_m\right\}\right] \text{,}
\end{equation}
i.e., there are two ADCs per receive antenna quantizing the real and imaginary part independently.

\section{LOS MIMO}

In LOS MIMO the channel matrix entries are modeled by spherical wave propagation \cite{Torkildson2011,Halsig2015}, i.e.,
\begin{equation}
(\matr{H})_{mn} = a_{mn}\cdot \exp\left(-j2\pi \frac{r_{mn}}{\lambda}\right) \text{,} \label{eq:hLOS}
\end{equation}
where $a_{mn}$ is the corresponding attenuation coefficient, $r_{mn}$ is the distance between transmit antenna $n$ and receive antenna $m$, and $\lambda$ is the wavelength of the carrier frequency. The value of $a_{mn}$ should in a LOS scenario be approximately equal across the different paths and can thus be neglected. Notice that we focus on pure LOS propagation here, which is a realistic assumption for, e.g., short-range indoor and highly directive backhaul outdoor links at millimeter wave frequencies.

\begin{figure}[!t]
\centering
\def\antenna{%
   -- +(0mm,-2.5mm) -- +(0mm,0.75mm) -- +(1.31mm,2.5mm) -- +(-1.31mm,2.5mm) -- +(0mm,0.75mm)
}

\definecolor{mycolor1}{rgb}{0.0784313753247261,0.168627455830574,0.549019634723663}%
\definecolor{mycolor2}{rgb}{1,1,1}%

\tikzset{>=latex}

\begin{tikzpicture}[font=\scriptsize]

\draw[color=mycolor1,fill=mycolor2,thick] (0,1) \antenna;
\draw[color=mycolor1,fill=mycolor2,thick] (0.75,1.5) \antenna;
\draw[fill=black] (1.5,2) circle (0.1mm);
\draw[fill=black] (1.4,1.95) circle (0.1mm);
\draw[fill=black] (1.6,2.05) circle (0.1mm);
\draw[color=mycolor1,fill=mycolor2,thick] (2.25,2.5) \antenna;

\node[font=\small] at (1,3.5) {$N$};

\node[font=\normalsize] at (3.5,1.75) {$\matr{H}$};

\begin{scope}[xshift=4.5cm]

\draw[color=mycolor1,fill=mycolor2,thick] (0,1) \antenna;
\draw[color=mycolor1,fill=mycolor2,thick] (0,0.75) -- (0.5,0.75);

\draw[color=mycolor1,fill=mycolor2,thick] (0.375,1.25) \antenna;
\draw[color=mycolor1,fill=mycolor2,thick] (0.375,1) -- (0.875,1);

\draw[color=mycolor1,fill=mycolor2,thick] (0.75,1.5) \antenna;
\draw[color=mycolor1,fill=mycolor2,thick] (0.75,1.25) -- (1.25,1.25);

\draw[color=mycolor1,fill=mycolor2,thick] (1.125,1.75) \antenna;
\draw[color=mycolor1,fill=mycolor2,thick] (1.125,1.5) -- (1.625,1.5);

\begin{scope}[xshift=0.5cm]
\draw[fill=black] (1.7,2) circle (0.1mm);
\draw[fill=black] (1.6,1.95) circle (0.1mm);
\draw[fill=black] (1.8,2.05) circle (0.1mm);
\end{scope}

\draw[color=mycolor1,fill=mycolor2,thick] (2.25,2.5) \antenna;
\draw[color=mycolor1,fill=mycolor2,thick] (2.25,2.25) -- (2.75,2.25);

\draw[color=mycolor1,fill=mycolor2,thick,double] (2.75,2) rectangle (3.5,2.5) node[pos=.5] {$Q_1[\cdot]$};
\draw[color=mycolor1,fill=mycolor2,thick,double] (1.625,1.25) rectangle (2.375,1.75) node[pos=.5] {$Q_1[\cdot]$};
\draw[color=mycolor1,fill=mycolor2,thick,double] (1.25,1) rectangle (2,1.5) node[pos=.5] {$Q_1[\cdot]$};
\draw[color=mycolor1,fill=mycolor2,thick,double] (0.875,0.75) rectangle (1.625,1.25) node[pos=.5] {$Q_1[\cdot]$};
\draw[color=mycolor1,fill=mycolor2,thick,double] (0.5,0.5) rectangle (1.25,1) node[pos=.5] {$Q_1[\cdot]$};

\node[font=\small] at (2.5,3.5) {$M=S\cdot N$};

\end{scope}

\end{tikzpicture}
\caption{Concept of spatial oversampling at the receiver for LOS MIMO systems.}
\label{fig:concept}
\end{figure}
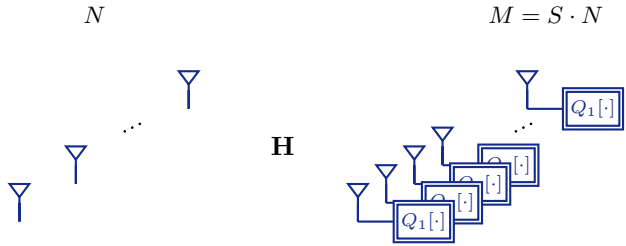

In LOS MIMO systems the spatial~degrees~of~freedom~(SDoF), i.e., number of spatially multiplexed streams, is determined by the size of the array \cite{Torkildson2011,Wang2014}. This is different from other channel types, in that the largest gain is given by the size of the arrays instead of the number of antennas. Hence, for a certain array size and transmission distance, we get the number of spatial streams that can be supported as well as their corresponding spatial sampling points, given that we have infinite resolution digital-to-analog~converters~(DACs) and ADCs. Note also that for most LOS MIMO systems the optimal antenna spacing is relatively large as compared to other systems.

\subsection{Spatial Oversampling}
We take a different approach here in that we assume the antenna outputs at the receiver are quantized with one bit. In that case, the full spatial multiplexing gain cannot be achieved, if we only sample at the optimal points. By using more spatial sampling points (antennas) than the optimal ones, each of them quantized with one bit, we hope to recover some of the information that is initially lost due to the quantization, comparable to the time domain oversampling gain shown in \cite{Krone2012,Halsig2014}. Due to the spherical wavefronts propagating across the receiver array, different spatial combinations of the transmit vectors are seen at different points in space, which can in certain scenarios generate sufficiently distinct spatial samples even when quantizing with one bit.

Similarly to an oversampling factor in the temporal domain, an oversampling factor can also be defined in the spatial domain. When oversampling in time, one takes a fixed number of additional samples per symbol interval compared to the number required for Nyquist rate sampling. An analog definition can be given for LOS MIMO when sampling additional points in space. If more antennas, than required for spatial multiplexing, are employed in the system, we speak of oversampling in space. To give an example, throughout this work we will assume a symmetric scenario, i.e., same array size at the transmitter and receiver. Then, the optimal sampling points and number of required antennas for spatial multiplexing at the transmitter and receiver are the same \cite{Torkildson2011,Halsig2015}. Given that there are $N$ spatial streams and transmit antennas, and $M$ receive antennas, the spatial sampling factor can be defined as
\begin{equation}
S = \frac{M}{N} \text{,}
\end{equation}
where we speak of spatial oversampling if $S>1$. This idea is illustrated in Fig.~\ref{fig:concept}. Note that for fair comparison, the array size should be kept constant when adding more antennas at the receiver, since increasing the overall array size (or aperture) always improves spatial multiplexing capabilities and thereby performance in the LOS case, as discussed earlier.



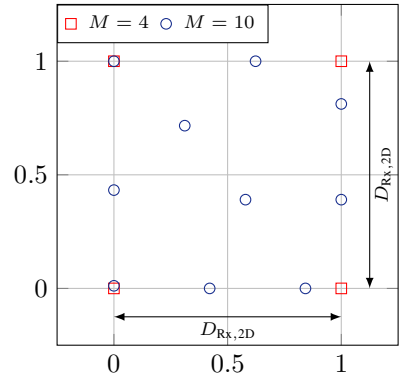
\begin{figure}[!t]
\centering
%
%
%
\definecolor{mycolor1}{rgb}{0.00000,0.44700,0.74100}%
\definecolor{mycolor2}{rgb}{0.85000,0.32500,0.09800}%

\definecolor{usualBlue}{rgb}{0.0784313753247261,0.168627455830574,0.549019634723663}

\begin{tikzpicture}

\begin{axis}[%
width=0.25\textwidth,
height=0.25\textwidth,
scale only axis,
xmin=-0.25,
xmax=1.25,
xmajorgrids,
ymin=-0.25,
ymax=1.25,
ymajorgrids,
legend style={at={(0.00,1.00)},anchor=north west,draw=black,fill=white,legend cell align=left,font=\scriptsize,column sep=2.5pt},
legend columns=2
]

\addplot [color=red,only marks,mark=square]
  table[row sep=crcr]{0	0\\
1	0\\
0	1\\
1	1\\
};
\addlegendentry{$M=4$};

\addplot [color=usualBlue,only marks,mark=o]
  table[row sep=crcr]{0.421123159552682	0\\
0.842402703536586	0\\
0	0.0114777466277879\\
0.578720456016097	0.390691113703279\\
1	0.390691113703279\\
0	0.432757290611691\\
0.311505026188565	0.716378645305846\\
1	0.811970657687182\\
0	1\\
0.623010052377129	1\\
};
\addlegendentry{$M=10$};

\draw[>=latex,<->] (0,-0.125) -- (1,-0.125) node[pos=0.5,font=\scriptsize,anchor=north,yshift=0.5mm,align=center] {$D_{\text{Rx},\text{2D}}$};
\draw[>=latex,<->] (1.125,0) -- (1.125,1) node[pos=0.5,font=\scriptsize,anchor=north,xshift=-0.5mm,rotate=90,align=center] {$D_{\text{Rx},\text{2D}}$};

\end{axis}
\end{tikzpicture}%
\caption{Best known two-dimensional antenna arrangements in a square with maximum minimum distance between the antennas for $M=4$ and $M=10$, according to \cite{Specht13}.}
\label{fig:circPack}
\end{figure}

\begin{figure*}[!t]
\centering
\subfloat[]{\columnplottwo
%
%
%
\definecolor{mycolor1}{rgb}{0.00000,0.44700,0.74100}%
\definecolor{mycolor2}{rgb}{0.85000,0.32500,0.09800}%
\definecolor{mycolor3}{rgb}{0.92900,0.69400,0.12500}%
\definecolor{mycolor4}{rgb}{0.49400,0.18400,0.55600}%
\definecolor{mycolor5}{rgb}{0.46600,0.67400,0.18800}%
\definecolor{mycolor6}{rgb}{0.30100,0.74500,0.93300}%

\definecolor{usualBlue}{rgb}{0.0784313753247261,0.168627455830574,0.549019634723663}

\begin{tikzpicture}

\begin{axis}[%
width=\figurewidth,
height=\figureheight,
scale only axis,
xmin=-5,
xmax=25,
xmajorgrids,
xlabel={SNR in \SI{}{\decibel}},
ymin=0,
ymax=5,
ymajorgrids,
ylabel={$I(\bm{X};\bm{Y}_Q)$ in \SI{}{bpcu}},
legend style={at={(0.00,1.00)},anchor=south west,draw=black,fill=white,legend cell align=left,font=\scriptsize,transpose legend},
legend columns=2
]
\addplot [color=black!50,dashed]
  table[row sep=crcr]{-5	0.792818322\\
-2	1.411438102\\
1	2.351273269\\
4	3.624492383\\
7	5.175628747\\
10	6.918863237\\
13	8.778117935\\
16	10.70175231\\
19	12.65942492\\
22	14.634632\\
25	16.61875048\\
};
\addlegendentry{Gaussian Input, $S=1$};

\addplot [color=black,dashdotted]	
  table[row sep=crcr]{-5	2.44210709808859\\
-2	3.33186181750047\\
1	3.87387262416289\\
4	3.99395717606787\\
7	3.99999903426751\\
10	4\\
13	4\\
16	4\\
19	4\\
22	4\\
25	4\\
};
\addlegendentry{$I(\bm{X};\bm{Y})$, $S=5$};

\addplot [color=usualBlue,solid]
  table[row sep=crcr]{-5	0.505130793\\
-2	0.881932313\\
1	1.372564126\\
4	1.80613484\\
7	1.982270176\\
10	1.999901578\\
13	2.000029239\\
16	2.000023178\\
19	2.000017053\\
22	2.000021831\\
25	2.00002026\\
};
\addlegendentry{$S=1$};

\addplot [color=usualBlue,solid,mark=x]
  table[row sep=crcr]{-5	0.903705593\\
-2	1.478653928\\
1	2.157084834\\
4	2.743036411\\
7	3.095443837\\
10	3.286511087\\
13	3.393771247\\
16	3.438589248\\
19	3.450382161\\
22	3.452960046\\
25	3.45313405\\
};
\addlegendentry{$S=2$};

\addplot [color=usualBlue,solid,mark=o]
  table[row sep=crcr]{-5	1.529062634\\
-2	2.298050151\\
1	3.041196678\\
4	3.507096699\\
7	3.638755928\\
10	3.599441872\\
13	3.529570869\\
16	3.479191127\\
19	3.45769493\\
22	3.452396978\\
25	3.452006147\\
};
\addlegendentry{$S=4$};

\addplot [color=usualBlue,solid,mark=square]
  table[row sep=crcr]{-5	1.782294071\\
-2	2.59224711\\
1	3.30090912\\
4	3.679624994\\
7	3.747925522\\
10	3.670767245\\
13	3.576792424\\
16	3.505398915\\
19	3.46775763\\
22	3.454834466\\
25	3.452777834\\
};
\addlegendentry{$S=5$};



\end{axis}
\end{tikzpicture}%
\label{fig:ULA_4QAM}}
\hfill
\subfloat[]{\columnplottwo
%
%
%
\definecolor{mycolor1}{rgb}{0.00000,0.44700,0.74100}%
\definecolor{mycolor2}{rgb}{0.85000,0.32500,0.09800}%
\definecolor{mycolor3}{rgb}{0.92900,0.69400,0.12500}%
\definecolor{mycolor4}{rgb}{0.49400,0.18400,0.55600}%
\definecolor{mycolor5}{rgb}{0.46600,0.67400,0.18800}%
\definecolor{mycolor6}{rgb}{0.30100,0.74500,0.93300}%

\definecolor{usualBlue}{rgb}{0.0784313753247261,0.168627455830574,0.549019634723663}

\begin{tikzpicture}

\begin{axis}[%
width=\figurewidth,
height=\figureheight,
scale only axis,
xmin=-5,
xmax=25,
xmajorgrids,
xlabel={SNR in \SI{}{\decibel}},
ymin=0,
ymax=9,
ymajorgrids,
ylabel={$I(\bm{X};\bm{Y}_Q)$ in \SI{}{bpcu}},
legend style={at={(0.00,1.00)},anchor=south west,draw=black,fill=white,legend cell align=left,font=\scriptsize,transpose legend},
legend columns=2	%
]
\addplot [color=black!50,dashed]
  table[row sep=crcr]{-5	0.792818322\\
-2	1.411438102\\
1	2.351273269\\
4	3.624492383\\
7	5.175628747\\
10	6.918863237\\
13	8.778117935\\
16	10.70175231\\
19	12.65942492\\
22	14.634632\\
25	16.61875048\\
};
\addlegendentry{Gaussian Input, $S=1$};

\addplot [color=black,dashdotted]	
  table[row sep=crcr]{-5	2.51480108710953\\
-2	3.69833204288507\\
1	5.06435792100744\\
4	6.49028906238312\\
7	7.58296971271539\\
10	7.96761373612153\\
13	7.99979356865234\\										
16	7.99999999999899\\
19	8\\
22	8\\
25	8\\
};
\addlegendentry{$I(\bm{X};\bm{Y})$, $S=5$};

\addplot [color=usualBlue,solid]
  table[row sep=crcr]{-5	0.496448229\\
-2	0.862388113\\
1	1.355369651\\
4	1.894471014\\
7	2.38119649\\
10	2.765143329\\
13	2.964454672\\
16	2.99965705\\
19	3.000334986\\
22	3.000330929\\
25	3.000332904\\
};
\addlegendentry{$S=1$};

\addplot [color=usualBlue,solid,mark=x]
  table[row sep=crcr]{-5	0.89693006\\
-2	1.461466996\\
1	2.152742484\\
4	2.847172787\\
7	3.4560822\\
10	3.951577592\\
13	4.28630726\\
16	4.465187463\\
19	4.586212752\\
22	4.679307949\\
25	4.744094887\\
};
\addlegendentry{$S=2$};

\addplot [color=usualBlue,solid,mark=o]
  table[row sep=crcr]{-5	1.538848255\\
-2	2.324587292\\
1	3.22029551\\
4	3.980527268\\
7	4.587798333\\
10	5.042774814\\
13	5.366170716\\
16	5.556100826\\
19	5.695844289\\
22	5.803403655\\
25	5.869352379\\
};
\addlegendentry{$S=4$};

\addplot [color=usualBlue,solid,mark=square]	
  table[row sep=crcr]{-5	1.78880819\\
-2	2.710011384\\
1	3.585700258\\
4	4.375641797\\
7	4.985291626\\
10	5.411754549\\
13	5.692227561\\
16	5.816203072\\
19	5.890035012\\
22	5.934846741\\
25	5.967014416\\
};
\addlegendentry{$S=5$};



\end{axis}
\end{tikzpicture}%
\label{fig:ULA_16QAM}}
\caption{Mutual information for $N=2$ input antennas in optimal arrangement, $M=SN$ receiving antennas, and one-dimensional transmit and receive arrays of sizes $D_{\text{Tx},\text{1D}}=D_{\text{Rx},\text{1D}}=\SI{0.5}{\meter}$: \protect\subref{fig:ULA_4QAM}~4-QAM; \protect\subref{fig:ULA_16QAM}~16-QAM.}
\label{fig:ULA_MI}
\end{figure*}

\subsection{Antenna Arrays} \label{sec:arrays}
The choice of sampling points, i.e., the antenna arrangement, is critical for LOS MIMO as discussed previously. It has been shown \cite{Bøhagen2006} that uniform~linear~arrays~(ULAs) and uniform~rectangular~arrays~(URAs) have the smallest size yielding the most SDoF with high resolution converters for the one- and two-dimensional case, respectively. On the transmitter side we will henceforth always choose the optimal ULA/URA according to the array size and SDoF number.

For the 1-bit quantized receiver array, we make the assumption that maximally separated spatial sampling points (antennas) will yield the best performance, as they give us the most distinct samples. For one-dimensional arrays, the antenna spacings $d_{\text{Rx},\text{1D}}$ and positions are then found from the receiver array size $D_{\text{Rx},\text{1D}}$ and the number of antennas $M$ as
\begin{equation}
d_{\text{Rx},\text{1D}} = \frac{D_{\text{Rx},\text{1D}}}{M-1} \text{.} \label{eq:ula}
\end{equation}
In the two dimensional case, where we wish to find the arrangement of points in a square with maximum minimum distance among them, this is known as the circle packing in a square problem. No general solution exists for an arbitrary number of $M$ points. However, some of the best known solutions, which are not necessarily optimal, can be found in \cite{Specht13}. The arrangements for $M=4$ and $M=10$ are given in Fig.~\ref{fig:circPack}.

%

\section{Mutual Information}
To evaluate the achievable throughput of the quantized system, we compute the mutual information $I(\cdot)$, assuming a memoryless channel, between the discrete input variable $\bm{X}$ and the discrete output variable $\bm{Y}_Q$, corresponding to the input and output symbol vectors of the system. In the general form it is given by
\begin{align}
I\left(\bm{X};\bm{Y}_Q\right) &= \sum_{\matr{x},\matr{y}_Q} p(\matr{x},\matr{y}_Q)\log_2\frac{p(\matr{x},\matr{y}_Q)}{p(\matr{x})p(\matr{y}_Q)} \\
&= H\left(\bm{X}\right)-H\left(\bm{X}|\bm{Y}_Q\right) \text{,} 
\end{align}
where $p(\cdot)$ denotes probability mass functions and $H(\cdot)$ denotes entropies. With the equivalencies 
\begin{equation}
p(\matr{x},\matr{y}_Q) = p(\matr{y}_Q|\matr{x})p(\matr{x}) = p(\matr{x}|\matr{y}_Q)p(\matr{y}_Q) \text{,}
\end{equation}
it suffices to compute the conditional probability $p(\matr{y}_Q|\matr{x})$. Starting point is a multivariate complex Gaussian distribution
\begin{equation}
p(\matr{y}|\matr{x}) = \frac{1}{\pi^M\det(\matr{N})}\exp\left(-(\matr{y}-\bm{\mu}_\matr{x})^{\hermit}\matr{N}^{-1}(\matr{y}-\bm{\mu}_\matr{x})\right)
\end{equation}
with covariance matrix and mean vector defined by
\begin{equation}
\matr{N} = \expec\left[\matr{n}\matr{n}^{\hermit}\right] = \sigma^2\matr{I}_{M} \hspace{1.0cm}\text{and}\hspace{1.0cm} \bm{\mu}_\matr{x} = \matr{H}\matr{x} \notag \text{.}
\end{equation}
Then, the conditional probabilities for the quantized output symbol vectors are gained by integrating 
\begin{equation}
p(\matr{y}_Q|\matr{x}) = \int\limits_{\matr{q}_l}^{\matr{q}_u} p(\matr{y}|\matr{x}) \diff\matr{y}
\end{equation}
where $\matr{q}_l$ and $\matr{q}_u$ are the complex integration limits corresponding to the output vector $\matr{y}_Q$, i.e.,
\begin{align}
{y}_{Q,m} \in \left\{
\begin{array}{ll}
\phantom{-}1+j: & {q}_{l,m}=[0,0] \text{, } {q}_{u,m}=[\infty,\infty] \\
\phantom{-}1-j: & {q}_{l,m}=[0,-\infty] \text{, } {q}_{u,m}=[\infty,0] \\
-1+j: & {q}_{l,m}=[-\infty,0] \text{, } {q}_{u,m}=[0,\infty] \\
-1-j: & {q}_{l,m}=[-\infty,-\infty] \text{, } {q}_{u,m}=[0,0]
\end{array}
\right\} \notag
\end{align}
with the first and second entry in ${q}_{l,m}$ and ${q}_{u,m}$ being the real and imaginary limit, respectively.


\begin{figure*}[!t]
\centering
\subfloat[]{\columnplottwo
%
%
%
\definecolor{mycolor1}{rgb}{0.00000,0.44700,0.74100}%
\definecolor{mycolor2}{rgb}{0.85000,0.32500,0.09800}%
\definecolor{mycolor3}{rgb}{0.92900,0.69400,0.12500}%
\definecolor{mycolor4}{rgb}{0.49400,0.18400,0.55600}%
\definecolor{mycolor5}{rgb}{0.46600,0.67400,0.18800}%
\definecolor{mycolor6}{rgb}{0.30100,0.74500,0.93300}%

\definecolor{usualBlue}{rgb}{0.0784313753247261,0.168627455830574,0.549019634723663}

\begin{tikzpicture}

\begin{axis}[%
width=\figurewidth,
height=\figureheight,
scale only axis,
xmin=-5,
xmax=25,
xmajorgrids,
xlabel={SNR in \SI{}{\decibel}},
ymin=0,
ymax=5,
ymajorgrids,
ylabel={$I(\bm{X};\bm{Y}_Q)$ in \SI{}{bpcu}},
legend style={at={(0.00,1.00)},anchor=south west,draw=black,fill=white,legend cell align=left,font=\scriptsize,transpose legend},
legend columns=2	%
]
\addplot [color=black!50,dashed]
  table[row sep=crcr]{-5	0.792818322\\
-2	1.411438102\\
1	2.351273269\\
4	3.624492382\\
7	5.175628747\\
10	6.918863237\\
13	8.778117934\\
16	10.70175231\\
19	12.65942492\\
22	14.634632\\
25	16.61875048\\
};
\addlegendentry{Gaussian Input, $S=1$};

\addplot [color=black,dashdotted]	
  table[row sep=crcr]{-5	2.48935311232338\\
-2	3.38611537483656\\
1	3.88935974388903\\
4	3.99773865440645\\
7	3.99999993392813\\	
10	4\\
13	4\\
16	4\\
19	4\\
22	4\\
25	4\\
};
\addlegendentry{$I(\bm{X};\bm{Y})$, $S=5$};

\addplot [color=usualBlue,solid]
  table[row sep=crcr]{-5	0.497909191\\
-2	0.865165955\\
1	1.356074076\\
4	1.865948882\\
7	2.279645176\\
10	2.579326804\\
13	2.755192639\\
16	2.807577856\\
19	2.811441064\\
22	2.811199727\\
25	2.811242434\\
};
\addlegendentry{$S=1$};

\addplot [color=usualBlue,solid,mark=x]
  table[row sep=crcr]{-5	0.919493099\\
-2	1.510264491\\
1	2.199359778\\
4	2.791964694\\
7	3.167134444\\
10	3.37848915\\
13	3.522647213\\
16	3.585960397\\
19	3.592073189\\
22	3.59204336\\
25	3.591128894\\
};
\addlegendentry{$S=2$};

\addplot [color=usualBlue,solid,mark=o]
  table[row sep=crcr]{-5	1.562040859\\
-2	2.353825593\\
1	3.110126951\\
4	3.612299161\\
7	3.849093125\\
10	3.941327497\\
13	3.966282629\\
16	3.969219007\\
19	3.969141183\\
22	3.968750558\\
25	3.968694118\\
};
\addlegendentry{$S=4$};

\addplot [color=usualBlue,solid,mark=square]
  table[row sep=crcr]{-5	1.800309136\\
-2	2.662545806\\
1	3.364962046\\
4	3.765382438\\
7	3.91761498\\
10	3.968608026\\
13	3.983183746\\
16	3.985701054\\
19	3.985975744\\
22	3.985905373\\
25	3.985885131\\
};
\addlegendentry{$S=5$};



\end{axis}
\end{tikzpicture}%
\label{fig:ULAura_4QAM}}
\hfill
\subfloat[]{\columnplottwo
%
%
%
\definecolor{mycolor1}{rgb}{0.00000,0.44700,0.74100}%
\definecolor{mycolor2}{rgb}{0.85000,0.32500,0.09800}%
\definecolor{mycolor3}{rgb}{0.92900,0.69400,0.12500}%
\definecolor{mycolor4}{rgb}{0.49400,0.18400,0.55600}%
\definecolor{mycolor5}{rgb}{0.46600,0.67400,0.18800}%
\definecolor{mycolor6}{rgb}{0.30100,0.74500,0.93300}%

\definecolor{usualBlue}{rgb}{0.0784313753247261,0.168627455830574,0.549019634723663}

\begin{tikzpicture}

\begin{axis}[%
width=\figurewidth,
height=\figureheight,
scale only axis,
xmin=-5,
xmax=25,
xmajorgrids,
xlabel={SNR in \SI{}{\decibel}},
ymin=0,
ymax=9,
ymajorgrids,
ylabel={$I(\bm{X};\bm{Y}_Q)$ in \SI{}{bpcu}},
legend style={at={(0.00,1.00)},anchor=south west,draw=black,fill=white,legend cell align=left,font=\scriptsize,transpose legend},
legend columns=2	%
]
\addplot [color=black!50,dashed]
  table[row sep=crcr]{-5	0.792818322\\
-2	1.411438102\\
1	2.351273269\\
4	3.624492382\\
7	5.175628747\\
10	6.918863237\\
13	8.778117934\\
16	10.70175231\\
19	12.65942492\\
22	14.634632\\
25	16.61875048\\
};
\addlegendentry{Gaussian Input, $S=1$};

\addplot [color=black,dashdotted]	
  table[row sep=crcr]{-5	2.59436037318854\\
-2	3.80671474133643\\
1	5.21192578970432\\
4	6.64135346923447\\
7	7.66253886927632\\
10	7.97888362046145\\
13	7.99991328839935\\
16	7.99999999999999\\										
19	8\\
22	8\\
25	8\\
};
\addlegendentry{$I(\bm{X};\bm{Y})$, $S=5$};

\addplot [color=usualBlue,solid]
  table[row sep=crcr]{-5	0.491764708\\
-2	0.850326192\\
1	1.335651321\\
4	1.8755209\\
7	2.375999566\\
10	2.792029228\\
13	3.108928716\\
16	3.336338959\\
19	3.49453607\\
22	3.593996313\\
25	3.658273444\\
};
\addlegendentry{$S=1$};

\addplot [color=usualBlue,solid,mark=x]
  table[row sep=crcr]{-5	0.913512331\\
-2	1.498081899\\
1	2.206568164\\
4	2.893024518\\
7	3.440935722\\
10	3.814257443\\
13	4.032039828\\
16	4.127465003\\
19	4.155265806\\
22	4.147349031\\
25	4.130911174\\
};
\addlegendentry{$S=2$};

\addplot [color=usualBlue,solid,mark=o]
  table[row sep=crcr]{-5	1.568434611\\
-2	2.392760304\\
1	3.298975082\\
4	4.14943175\\
7	4.867079419\\
10	5.431057674\\
13	5.837968003\\
16	6.077839077\\
19	6.192667257\\
22	6.231937051\\
25	6.242974231\\
};
\addlegendentry{$S=4$};

\addplot [color=usualBlue,solid,mark=square]
  table[row sep=crcr]{-5	1.904679972\\
-2	2.775142475\\
1	3.699718358\\
4	4.544915425\\
7	5.237157992\\
10	5.752145598\\
13	6.106778537\\
16	6.305483933\\
19	6.452229436\\
22	6.562382324\\
25	6.624855629\\
};
\addlegendentry{$S=5$};



\end{axis}
\end{tikzpicture}%
\label{fig:ULAura_16QAM}}
\caption{Mutual information for $N=2$ input antennas in optimal arrangement, $M=SN$ receiving antennas, and one-dimensional transmit array, two-dimensional receive array, of sizes $D_{\text{Tx},\text{1D}}=D_{\text{Rx},\text{2D}}=\SI{0.5}{\meter}$: \protect\subref{fig:ULAura_4QAM}~4-QAM; \protect\subref{fig:ULAura_16QAM}~16-QAM.}
\label{fig:ULAura_MI}
\end{figure*}

\subsection{High SNR Regime}
When $\sigma\to 0$, i.e., the high SNR regime, the integration reduces to
\begin{equation}
p(\matr{y}_Q|\matr{x}) = \int\limits_{\matr{q}_l}^{\matr{q}_u} \delta(\matr{y}-\bm{\mu}_\matr{x}) \diff\matr{y} \label{eq:int_highsnr}
\end{equation}
where $\delta(\cdot)$ is the multidimensional Dirac delta distribution, and the boundary case is defined as 
\begin{align}
p(y_{Q,m}=-1|\mu_{\matr{x},m}=0)&=	p(y_{Q,m}=1|\mu_{\matr{x},m}=0) \\
\int\limits_{-\infty}^{0} \delta(y_m) \diff y_m &= \int\limits_{0}^{\infty} \delta(y_m) \diff y_m = \frac{1}{2}
\end{align}
assuming that the quantizers will decide for each of the two possible values the same amount of the time.

For a transmit vector design exploiting the results shown further on and improving on them, which will not be part of this work, a unique matching between $\bm{\mu}_\matr{x}$ and $\matr{y}_Q$ in \eqref{eq:int_highsnr} is desired. Transmit vectors generating, e.g., an all zero vector in $\bm{\mu}_\matr{x}$, should thus be avoided as they cannot be uniquely decoded without additional effort.

\section{Numerical Results}
For the following results we consider two complex input alphabets, 4-QAM and 16-QAM, and investigate their performance for different system setups. The array sizes are fixed to $D_{\text{Tx},\text{1D}}=D_{\text{Rx},\text{1D}}=\SI{0.5}{\meter}$ and $D_{\text{Tx},\text{2D}}=D_{\text{Rx},\text{2D}}=\SI{0.5}{\meter}$, the transmission distance is \SI{100}{\meter} and the wavelength is $\lambda=\SI{5}{\milli\meter}$. Furthermore, the broadsides of transmit and receive array are always facing each other, i.e., perfect alignment no rotations or translations are considered. For comparison we have added, where applicable, the unquantized channel capacity with Gaussian input for the standard case of $M=N$ or, equivalently, $S=1$ and the unquantized mutual information $I(\bm{X};\bm{Y})$ for the corresponding input and the highest spatial sampling factor, which is for the considered cases at high SNRs equal to the source entropy $H(\bm{X})$, i.e., the highest achievable rate with this input alphabet and number of antennas. Finally, under a transmit power constraint, i.e., $\expec\left[\matr{x}\matr{x}^{\hermit}\right] = \frac{1}{N}\matr{I}_{N}$, the signal-to-noise ratio is defined as $\text{SNR} = 10\log_{10}(\frac{1}{\sigma^2})$.


\subsection{Uniform one-dimensional Antenna Arrays}
In Fig.~\ref{fig:ULA_MI} we show the mutual information results in bits~per~channel~use~(\SI{}{bpcu}) for the two input alphabets with an optimally arranged transmit ULA with $N=2$ for different spatial sampling factors $S$, where the receiver array is spaced according to equation \eqref{eq:ula}. It is visible that by increasing $S$ the mutual information increases. At high SNRs there seems to exist a limit beyond which the rate cannot be improved, even when oversampling more. This is likely due to the fact that either the receive array is not optimized for the given situation or that the given channel cannot generate sufficiently different spatial sampling points. Generally, the higher the dimensionality of the system, i.e., number of transmit antennas and cardinality of input alphabet, the higher the required resolution or, equivalently, the more helpful oversampling becomes.

A parallel to time domain oversampling can be drawn when looking at the $\text{SNR}$ region between \SI{0}{\decibel} and \SI{10}{\decibel} in Fig.~\ref{fig:ULA_4QAM}. In that region, the same effect of a stochastic resonance \cite{Krone2012}, whereby the channel noise effectively acts as a dithering signal and thus improves the mutual information, occurs. We conjecture that as $S\to\infty$ the full rate available from the source is achieved for that region, i.e., $I(\bm{X};\bm{Y}_Q)\to H(\bm{X})$, as is the case for the time domain process.

\subsection{Two-dimensional Antenna Arrays}
Fig.~\ref{fig:ULAura_MI} shows the case where the one-dimensional array at the transmitter from above is kept and a two-dimensional array is used at the receiver, in order to evaluate the benefit of the second dimension. The receiver array is chosen according to \cite{Specht13} as explained in section \ref{sec:arrays}. It can be seen that compared to Fig.~\ref{fig:ULA_MI}, the second dimension improves the performance in all cases significantly. For 4-QAM it is possible to achieve the full rate of \SI{4}{bpcu} that is available from the transmitter input with $S=4$ and $S=5$. Again, for the higher cardinality alphabet of 16-QAM the overall rate is improved, but there is a gap between source entropy $H(\bm{X})$ and mutual information $I(\bm{X};\bm{Y}_Q)$ at high SNRs.

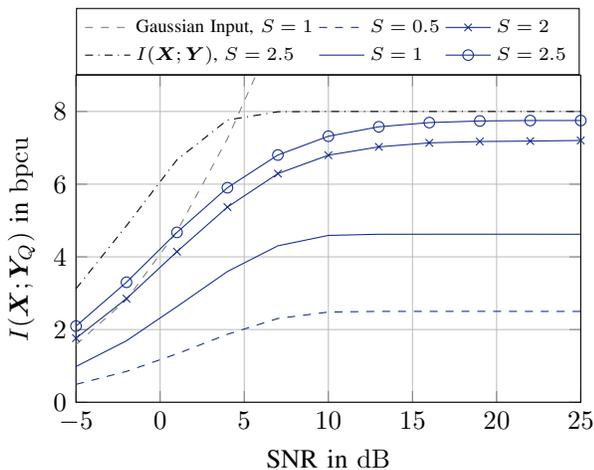
\begin{figure}[!t]
\centering
\columnplot
%
%
%
\definecolor{mycolor1}{rgb}{0.00000,0.44700,0.74100}%
\definecolor{mycolor2}{rgb}{0.85000,0.32500,0.09800}%
\definecolor{mycolor3}{rgb}{0.92900,0.69400,0.12500}%
\definecolor{mycolor4}{rgb}{0.49400,0.18400,0.55600}%
\definecolor{mycolor5}{rgb}{0.46600,0.67400,0.18800}%
\definecolor{mycolor6}{rgb}{0.30100,0.74500,0.93300}%

\definecolor{usualBlue}{rgb}{0.0784313753247261,0.168627455830574,0.549019634723663}

\begin{tikzpicture}

\begin{axis}[%
width=\figurewidth,
height=\figureheight,
scale only axis,
xmin=-5,
xmax=25,
xmajorgrids,
xlabel={SNR in \SI{}{\decibel}},
ymin=0,
ymax=9,
ymajorgrids,
ylabel={$I(\bm{X};\bm{Y}_Q)$ in \SI{}{bpcu}},
legend style={at={(0.00,1.00)},anchor=south west,draw=black,fill=white,legend cell align=left,font=\scriptsize,transpose legend},
legend columns=2	%
]
\addplot [color=black!50,dashed]
  table[row sep=crcr]{-5	1.585636645\\
-2	2.822876203\\
1	4.702546538\\
4	7.248984764\\
7	10.35125749\\
10	13.83772647\\
13	17.55623587\\
16	21.40350461\\
19	25.31884984\\
22	29.26926401\\
25	33.23750096\\
};
\addlegendentry{Gaussian Input, $S=1$};

\addplot [color=black,dashdotted]	
  table[row sep=crcr]{-5	3.12335308063945\\
-2	4.85728794114735\\
1	6.67015102700598\\
4	7.76482257934890\\
7	7.99241737765170\\
10	7.99999427222192\\
13	8\\
16	8\\
19	8\\
22	8\\
25	8\\
};
\addlegendentry{$I(\bm{X};\bm{Y})$, $S=2.5$};

\addplot [color=usualBlue,dashed]
  table[row sep=crcr]{-5	0.491800133\\
-2	0.848368161\\
1	1.33215595\\
4	1.869098564\\
7	2.304672729\\
10	2.481526766\\
13	2.499774271\\
16	2.499709161\\
19	2.501275369\\
22	2.499551206\\
25	2.498952439\\
};
\addlegendentry{$S=0.5$};

\addplot [color=usualBlue,solid]
  table[row sep=crcr]{-5	0.982297443\\
-2	1.691719461\\
1	2.631839087\\
4	3.597038047\\
7	4.304961365\\
10	4.590666774\\
13	4.622637655\\
16	4.623139997\\
19	4.623267668\\
22	4.623244008\\
25	4.622541425\\
};
\addlegendentry{$S=1$};

\addplot [color=usualBlue,solid,mark=x]
  table[row sep=crcr]{-5	1.762822773\\
-2	2.850701148\\
1	4.144017989\\
4	5.369964621\\
7	6.291235646\\
10	6.797233503\\
13	7.026482667\\
16	7.13713934\\
19	7.173729837\\
22	7.186182637\\
25	7.202577392\\
};
\addlegendentry{$S=2$};

\addplot [color=usualBlue,solid,mark=o]
  table[row sep=crcr]{-5	2.094966349\\
-2	3.301836936\\
1	4.673475997\\
4	5.903124589\\
7	6.802168535\\
10	7.319186231\\
13	7.578856972\\
16	7.694623761\\
19	7.739048485\\
22	7.751068895\\
25	7.751487588\\
};
\addlegendentry{$S=2.5$};




\end{axis}
\end{tikzpicture}%
\caption{Mutual information for 4-QAM, $N=4$ input antennas in optimal arrangement, $M=SN$ receiving antennas, and two-dimensional transmit and receive arrays of sizes $D_{\text{Tx},\text{2D}}=D_{\text{Rx},\text{2D}}=\SI{0.5}{\meter}$.}
\label{fig:URA_4QAM}
\end{figure}

In Fig.~\ref{fig:URA_4QAM} we have computed the results for two-dimensional arrays on both sides of the link. At the transmitter we use an optimally arranged URA with $N=4$, at the receiver the arrays are again arranged according to \cite{Specht13}. Similar to the other results, the performance improves when $S$ increases. When comparing the results with the ones in Fig.~\ref{fig:ULAura_MI}, one can see how the performance changes for different transmitter arrangements, as they both use the same numbers and structures for the receive antenna arrays. For example, comparing the results for $S=2.5$ with the results in Fig.~\ref{fig:ULAura_16QAM} for $S=5$, both yielding $M=10$, shows that the two-dimensional transmit array with $N=4$ is preferable in this case as its mutual information is significantly higher, and almost achieves the available rate from the source, which is for both cases \SI{8}{bpcu}. Note that this comes at the cost of an increased largest array dimension at the transmitter.

\begin{figure}[!t]
\centering
\columnplot
%
%
%
\definecolor{mycolor1}{rgb}{0.00000,0.44700,0.74100}%
\definecolor{mycolor2}{rgb}{0.85000,0.32500,0.09800}%
\definecolor{mycolor3}{rgb}{0.92900,0.69400,0.12500}%
\definecolor{mycolor4}{rgb}{0.49400,0.18400,0.55600}%
\definecolor{usualBlue}{rgb}{0.0784313753247261,0.168627455830574,0.549019634723663}

\begin{tikzpicture}

\begin{axis}[%
width=\figurewidth,
height=\figureheight,
scale only axis,
xmin=0.5,
xmax=5,
xmajorgrids,
xlabel={Spatial Sampling Factor $S$},
ymin=0,
ymax=8,
ymajorgrids,
ylabel={$I(\bm{X};\bm{Y}_Q)$ in \SI{}{bpcu}},
legend style={at={(1.00,0.00)},anchor=south east,draw=black,fill=white,legend cell align=left,font=\scriptsize},
legend columns=2	%
]

%
%
%

\addplot [color=usualBlue,solid]
  table[row sep=crcr]{0.5	1.500109773\\									
1	1.62255175\\
1.5	2.74168432\\
2	3.560755937\\
2.5	3.706436461\\
3	3.99848396\\
3.5	3.999283771\\
4	3.99984626\\
4.5	3.999859969\\
5	3.999913725\\
};
\addlegendentry{$\text{1D}\times\text{1D}$, 4-QAM};

\addplot [color=red,solid]
  table[row sep=crcr]{0.5	1.784563963\\
1	2.087041537\\
1.5	3.713557058\\
2	4.736803925\\
2.5	5.366122836\\
3	6.003490533\\
3.5	6.263417091\\
4	6.672795854\\
4.5	6.808493311\\
5	7.063675674\\
};
\addlegendentry{$\text{1D}\times\text{1D}$, 16-QAM};

\addplot [color=usualBlue,solid,mark=x]
  table[row sep=crcr]{0.5	1.500109773\\ 									
1	3.209928371\\
1.5	3.766433021\\
2	3.7243737\\
2.5	3.846097518\\
3	3.746516034\\
3.5	3.88442253\\
4	3.939048634\\
4.5	3.915183872\\
5	3.952698444\\
};
\addlegendentry{$\text{1D}\times\text{2D}$, 4-QAM};

\addplot [color=red,solid,mark=x]
  table[row sep=crcr]{0.5	1.784563963\\ 				
1	3.405018246\\
1.5	4.6027856\\
2	3.997427808\\
2.5	5.235013434\\
3	5.996256232\\
3.5	6.419249482\\
4	6.469493572\\
4.5	6.472721083\\
5	6.848188724\\
};
\addlegendentry{$\text{1D}\times\text{2D}$, 16-QAM};

\addplot [color=usualBlue,solid,mark=o]
  table[row sep=crcr]{0.25	1.453279566\\
0.5	2.500399343\\
0.75	4.574872472\\
1	4.622773998\\
1.25	5.489129382\\
1.5	6.853378823\\
1.75	7.284675125\\
2	7.177692983\\
2.25	7.569141528\\
2.5	7.745058164\\
};
\addlegendentry{$\text{2D}\times\text{2D}$, 4-QAM};

\end{axis}
\end{tikzpicture}%
\caption{Mutual information for the different input alphabets and antenna arrangements, array sizes of $D_{\text{Tx},\text{1D}}=D_{\text{Rx},\text{1D}}=1/\sqrt{2}\hspace{0.5mm}\SI{}{\meter}$, $D_{\text{Tx},\text{2D}}=D_{\text{Rx},\text{2D}}=\SI{0.5}{\meter}$, and $\text{SNR}=\SI{20}{\decibel}$.}
\label{fig:Oversamp}
\end{figure}
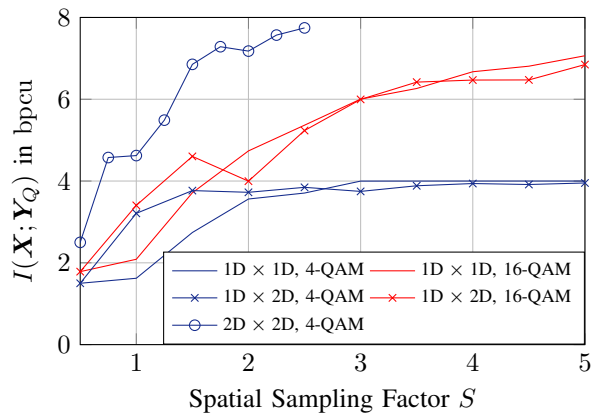

\subsection{Spatial Sampling Factor}
Finally, we show the performance for all of the considered cases so far versus the spatial sampling factor $S$ at $\text{SNR}=\SI{20}{\decibel}$ in Fig.~\ref{fig:Oversamp}. Note that for a fair comparison we have fixed the largest array dimension for all cases to $1/\sqrt{2}\hspace{0.5mm}\SI{}{\meter}$, i.e., the one-dimensional ULAs are slightly larger than in the previous cases. As for the preceding results, in the one dimensional transmit array case $N=2$ while in the two dimensional transmit array case $N=4$. Significant gains can be achieved in all cases when increasing $S$. Furthermore, a difference between one- and two-dimensional arrays is visible. While for large values of $S$ the performance is very similar when comparing one- and two-dimensional receive arrays, for low values of $S$ the two-dimensional case has superior performance. The reason for this could be, that for lower oversampling factors the actual position of the spatial sampling points has a bigger impact, which was not optimized in this work, or that the two-dimensional case offers inherently more diverse samples.

The non-continuous behavior of the 2D cases can similarly be explained by the receiver array assumption for these cases. We assumed that maximally separated antennas would yield the best results, because the spatially sampled points are maximally different. From the results in Fig.~\ref{fig:Oversamp} we infer, however, that for certain numbers of receive antennas, e.g., $M=4$ and $M=8$, there should be other antenna arrangements that have a better performance and generate a more continuous curve, as is the case for the one-dimensional arrays.

\section{Conclusion}
In this paper we have discussed the applicability of spatial oversampling at the receiver for 1-bit quantized LOS MIMO systems. Receiver antenna arrangements performing the sampling in space were introduced for the one-dimensional and two-dimensional case, based on the idea that maximally separated antennas gather the most distinct spatial points. Generally, the results show that spatial oversampling helps to increase the rate, just as oversampling in time does, and that the two-dimensional case is preferable. The results also reveal that the chosen receive antenna arrangements in the two-dimensional case are not optimal for all antenna numbers and can possibly be optimized. Furthermore, it is shown that for certain transmission scenarios the full available rate from the input alphabet can be achieved, thus fully compensating the loss that is incurred due to the quantization by spatially oversampling the received signal. Hence, we conclude that it is possible, to some extent, to trade off amplitude resolution of the quantizer against number of antennas in LOS MIMO systems.

\bibliographystyle{IEEEtran}
\bibliography{references}

\begin{thebibliography}{10}
\providecommand{\url}[1]{#1}
\csname url@samestyle\endcsname
\providecommand{\newblock}{\relax}
\providecommand{\bibinfo}[2]{#2}
\providecommand{\BIBentrySTDinterwordspacing}{\spaceskip=0pt\relax}
\providecommand{\BIBentryALTinterwordstretchfactor}{4}
\providecommand{\BIBentryALTinterwordspacing}{\spaceskip=\fontdimen2\font plus
\BIBentryALTinterwordstretchfactor\fontdimen3\font minus
  \fontdimen4\font\relax}
\providecommand{\BIBforeignlanguage}[2]{{%
\expandafter\ifx\csname l@#1\endcsname\relax
\typeout{** WARNING: IEEEtran.bst: No hyphenation pattern has been}%
\typeout{** loaded for the language `#1'. Using the pattern for}%
\typeout{** the default language instead.}%
\else
\language=\csname l@#1\endcsname
\fi
#2}}
\providecommand{\BIBdecl}{\relax}
\BIBdecl

\bibitem{Song2016}
X.~Song, T.~H{\"{a}}lsig, W.~Rave, B.~Lankl, and G.~Fettweis, ``{Analog
  Equalization and Low Resolution Quantization in Strong Line-of-Sight MIMO
  Communication},'' in \emph{Proc. IEEE Int. Conf. Commun.}, 2016, pp. 1--7.

\bibitem{Torkildson2011}
E.~Torkildson, U.~Madhow, and M.~Rodwell, ``{Indoor Millimeter Wave MIMO:
  Feasibility and Performance},'' \emph{IEEE Trans. Wirel. Commun.}, vol.~10,
  no.~12, pp. 4150--4160, 2011.

\bibitem{Halsig2015}
T.~H{\"{a}}lsig and B.~Lankl, ``{Array Size Reduction for High-Rank LOS MIMO
  ULAs},'' \emph{IEEE Wirel. Commun. Lett.}, vol.~4, no.~6, pp. 649--652, 2015.

\bibitem{Krone2012}
S.~Krone and G.~Fettweis, ``{Communications with 1-Bit Quantization and
  Oversampling at the Receiver: Benefiting from Inter-Symbol-Interference},''
  in \emph{Proc. IEEE 23rd Int. Symp. Pers. Indoor Mob. Radio Commun.}, 2012,
  pp. 2408--2413.

\bibitem{Halsig2014}
T.~H{\"{a}}lsig, L.~Landau, and G.~Fettweis, ``{Spectral Efficient
  Communications employing 1-Bit Quantization and Oversampling at the
  Receiver},'' in \emph{Proc. IEEE 80th Veh. Technol. Conf.}, 2014, pp. 1--5.

\bibitem{Landau2015}
L.~Landau, M.~D{\"{o}}rpinghaus, and G.~Fettweis, ``{Communication Employing
  1-Bit Quantization and Oversampling at the Receiver: Faster-than-Nyquist
  Signaling and Sequence Design},'' in \emph{Proc. IEEE Int. Conf. Ubiqitous
  Wirel. Broadband}, 2015, pp. 1--5.

\bibitem{Mezghani2007}
A.~Mezghani and J.~A. Nossek, ``{On Ultra-Wideband MIMO Systems with 1-bit
  Quantized Outputs: Performance Analysis and Input Optimization},'' in
  \emph{Proc. IEEE Int. Symp. Inf. Theory}, 2007, pp. 1286--1289.

\bibitem{Mo2015}
J.~Mo and R.~W. Heath, ``{Capacity Analysis of One-Bit Quantized MIMO Systems
  With Transmitter Channel State Information},'' \emph{IEEE Trans. Signal
  Process.}, vol.~63, no.~20, pp. 5498--5512, 2015.

\bibitem{Corey2016}
R.~M. Corey and A.~C. Singer, ``{Spatial Sigma-Delta Signal Acquisition for
  Wideband Beamforming Arrays},'' in \emph{Proc. Int. ITG Work. Smart
  Antennas}, 2016, pp. 524--530.

\bibitem{Palguna2016}
D.~S. Palguna, D.~J. Love, T.~A. Thomas, and A.~Gosh, ``{Millimeter Wave
  Receiver Design Using Parallel Delta Sigma ADCS and Low Precision
  Quantization},'' \emph{IEEE Trans. Wirel. Commun.}, vol.~15, no.~10, pp.
  6556--6569, 2016.

\bibitem{Ugurlu2016}
U.~Ugurlu and R.~Wichman, ``{Enabling Low-Resolution ADC with High-Order
  Modulations for Millimeter-Wave Systems},'' in \emph{Proc. IEEE Int. Conf.
  Commun.}, 2016, pp. 1--6.

\bibitem{Jacobsson2015}
S.~Jacobsson, G.~Durisi, M.~Coldrey, U.~Gustavsson, and C.~Studer, ``{One-Bit
  Massive MIMO: Channel Estimation and High-Order Modulations},'' in
  \emph{Proc. IEEE Int. Conf. Commun.}, 2015, pp. 1304--1309.

\bibitem{Wang2014}
P.~Wang, Y.~Li, X.~Yuan, L.~Song, and B.~Vucetic, ``{Tens of Gigabits Wireless
  Communications over E-band LoS MIMO Channels with Uniform Linear Antenna
  Arrays},'' \emph{IEEE Trans. Wirel. Commun.}, vol.~13, no.~7, pp. 3791--3805,
  2014.

\bibitem{Specht13}
\BIBentryALTinterwordspacing
E.~Specht. (2013) {The best known packings of equal circles in a square (up to
  $N = 10000$)}. [Online]. Available:
  \url{\url{http://hydra.nat.uni-magdeburg.de/packing/csq/csq.html}}
\BIBentrySTDinterwordspacing

\bibitem{Bøhagen2006}
F.~B{\o}hagen, P.~Orten, and G.~E. {\O}ien, ``{Optimal Design of Uniform Planar
  Antenna Arrays for Strong Line-of-Sight MIMO channels},'' in \emph{Proc. IEEE
  7th Work. Signal Process. Adv. Wirel. Commun.}, 2006, pp. 1 -- 5.

\end{thebibliography}
%

\end{document}